\documentclass[letterpaper,12pt,pra]{revtex4}
\usepackage{epsfig,amsfonts,amsmath}
\usepackage{graphics,times,hyperref,fullpage}
\usepackage{setspace} 
\usepackage[margin=2.6cm]{geometry}
\newcommand{\be}{\begin{equation}}
\newcommand{\ee}{\end{equation}}
\newcommand{\ba}{\begin{eqnarray}}
\newcommand{\ea}{\end{eqnarray}}

\def\bea{\begin{eqnarray}}
\def\eea{\end{eqnarray}}
\def\ben{\begin{eqnarray*}}
\def\een{\end{eqnarray*}}

\def\>{\rangle}
\def\<{\langle}

\newcommand{\eq}[1]{Eq.~(\ref{eq:#1})}
\newcommand{\fig}[1]{Fig.~\ref{fig:#1}}

\begin{document}

\title{Experimental demonstration of a surface-electrode multipole ion trap}
\author{Mark Maurice, Curtis Allen, Dylan Green, Andrew Farr, Timothy Burke, Russell Hilleke, and Robert Clark}

% \thanks{robclark@physics.utexas.edu}}
%\institute{Department of Physics, The Citadel, Charleston, SC 29409}
\affiliation{Department of Physics, The Citadel, Charleston, SC 29409}
\date{\today}

\begin{abstract}
We report on the design and experimental characterization of a surface-electrode multipole ion trap. Individual microscopic sugar particles are confined in the trap. The trajectories of driven particle motion are compared with a theoretical model, both to verify qualitative predictions of the model, and to measure the charge-to-mass ratio of the confined particle. The generation of harmonics of the driving frequency is observed as a key signature of the nonlinear nature of the trap. We remark on possible applications of our traps, including to mass spectrometry. 

\end{abstract} 

\maketitle

\section{Introduction}

Recent years have seen a wealth of innovation in the design of radiofrequency (RF) ion traps \cite{Madsen:04,Chiaverini:05,Monroe:06,Pearson:06,Seidelin:06,smit:09,Clark:09,THKim:10,Moehring:11,Shu:14}. To some extent, this has been driven by the utility of microfabricated ion traps for quantum information processing (QIP). However, there is still active interest in the use of RF ion traps in atomic and molecular spectroscopy \cite{Schmidt:05,Rosenband:08,Chou:11,CRClark:10,Rugango:15}, mass spectrometry \cite{Lorenz:11}, nonlinear dynamics \cite{Akerman:10,Wu:14,Vinitsky:15}, and other areas. Although most work geared toward QIP has focused on quadrupole ion traps, which have an electric potential and time-averaged pseudopotential that are quadratic near the trap center, there are also groups pursuing new designs and applications of multipole ion traps  \cite{Debatin:08,Champenois:09,Champenois:10,Marciante:11,Lorenz:11}. Our working definition of a multipole trap is that it is a trap that features a potential landscape that is of a higher order (going as $x^4$ or $x^6$, for example, where $x$ is the position of a trapped ion with respect to the trap center). The recent growth of interest in nonlinear dynamics using ions, in particular, has involved the use of quadrupole ion traps, but within which the ion is excited to a high enough energy that it is acted on by higher-order terms in the trapping potential \cite{Akerman:10,Wu:14,Vinitsky:15}. For such experiments, a multipole trap would be advantageous. 

We have recently proposed a trap design for a surface-electrode multipole ion trap (SEMT) \cite{Clark:13}. There are a number of reasons why this design might be of interest to researchers. Firstly, the trapping potential is highly nonlinear near the trap center, which eliminates the need to impart as much energy to the ion to see nonlinear behavior, if that is one's goal. Secondly, no DC voltages are required for ion confinement, as is the case with linear multipole traps. This results in less micromotion than one would otherwise encounter, which is beneficial in precision laser spectroscopy and other applications. (Micromotion is driven motion at the RF frequency which is present whenever the ion is located anywhere the RF driving field does not vanish. In linear multipole traps, the DC voltages on the endcaps result in electric fields that push ions further from the RF null than they would otherwise be \cite{Champenois:10}.) Thirdly, the designs set forth in Ref.~\cite{Clark:13} are already being used, in a modified form, in the design of ion traps which will be at the heart of exotic experiments on time crystals \cite{Li:12,Wang:15}. Fourthly, ideas for new applications continue to arise, including the possibility of using SEMT's as the basis of a new method of mass spectrometry. This will be discussed at the conclusion of this article. 

In this article, we describe the design and experimental characterization of a SEMT. Our key result is the demonstration of frequency upconversion in the trajectory of an ion which is close to the trap center, which is a hallmark of nonlinear behavior. In Sec.~II, we describe the mathematical model of the trap. In Sec.~III, we present our experimental setup and main results. Sec.~IV contains a discussion of our results, and some reflections on possible applications of this kind of trap. 

\section{Trap design and modeling}

\begin{figure}
\begin{center}
\includegraphics[width=.5\textwidth]{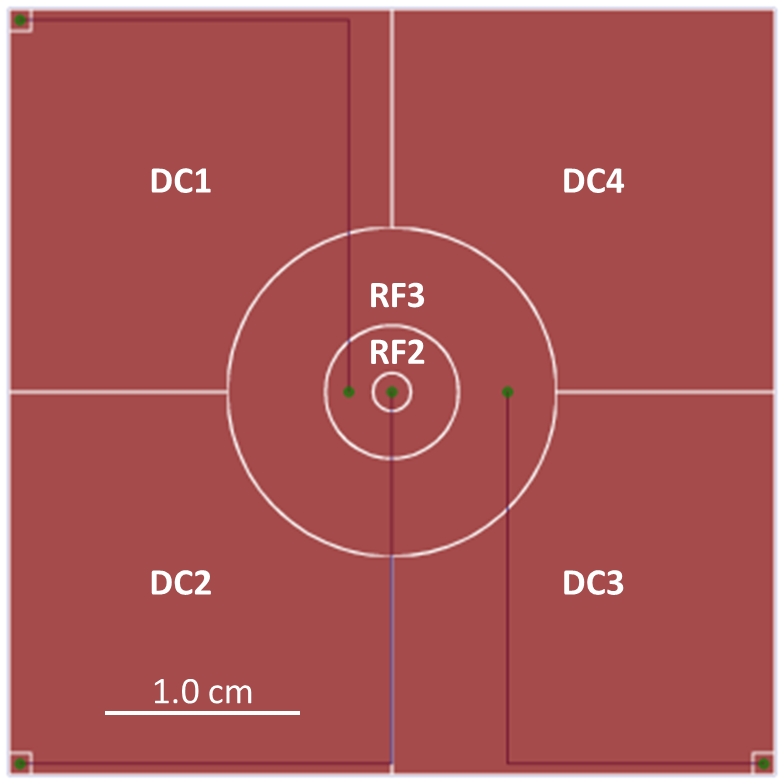}
\end{center}
\caption{Image of the trap as rendered by the EAGLE program. The overall width of the trap is 40.2~mm. The electrodes which may be set to a static voltage are labeled as ``DC1'' through ``DC4.'' The electrodes to which an oscillating voltage may be applied are labeled ``RF2'' and ``RF3.'' The center electrode is ``RF1,'' although it is not labeled. The nominal radii of the RF electrodes are 1.02~mm, 3.48~mm, and 8.62~mm. There is 0.15~mm gap between these electrodes. }\label{fig:fig1}
\end{figure}

The design of our SEMT is presented in \fig{fig1}. The trap was designed using the EAGLE PCB layout software and manufactured by Sunstone Circuits. It employs a silver finish to avoid corrosion and to make the electrodes equipotential surfaces to the greatest extent possible. The electrode radii chosen are only one of an infinite number of possibilities that will generate a multipole trap. These particular values were chosen because they roughly maximize the expected trap depth for a given RF voltage and frequency, and for an ion charge-to-mass ratio of 1~mC/kg (an order of magnitude estimate for dust particles). A surface-electrode multipole trap is created by applying alternating voltage to RF1 and RF3 while grounding RF2. 

The time-averaged effective potential which governs the motion of the ions, called the \emph{pseudopotential} and denoted $\Psi(r,z)$, is given by 

\begin{equation}
\Psi(r,z) = \frac{Q^2}{4 m \Omega^2} \left | \nabla \Phi(r,z) \right| ^2, 
\end{equation}

\noindent where $Q$ is the charge of the ion, $m$ is the mass of the ion, 
$\Omega$ is the (angular) frequency at which the trap is driven, and $\nabla \Phi(r,z)$ is the RF field amplitude. We define the trap axis as the line perpendicular to the plane of the trap and intersecting the center of the trap electrodes. The $z$ coordinate measures the displacement above the electrodes, measured along the trap axis, and the coordinate $r$ measures the displacement perpendicular to the trap axis in a plane parallel to the trap electrodes. The calculation of $\Phi(r,z)$ for many trap geometries, including ring traps, is a problem in electrostatics to which several authors have recently made important contributions \cite{Wesenberg:08,House:08,Schmied:09,Schmied:10}. In the present work, we use the methods outlined by Schmied \cite{Schmied:10}. The purpose of using these methods is to take into account the effect of gaps between the electrodes, which were set to be the minimum allowed by the manufacturer (Sunstone Circuits). However, we do note that the true potential is indeed very close (within 1\% inside the region of interest) to that obtained from the gapless approximation. In the gapless approximation, the electric potential is given by 

\begin{equation}
\Phi(z,r) = \int_0^{\infty} J_0(kr) e^{-kz} A_0(k) dk  
\label{eq:phi} 
\end{equation} 

\noindent where $J_0$ is the Bessel function of zeroth order and $A_0(k) = \sum_{i=1}^N A_i(k)$. The $A_i$ are given by 

\begin{equation}
A_i(k) = V_i \left ( r_{i+1} J_1(kr_{i+1}) - r_i J_1(kr_i) \right ) ,
\end{equation} 

\noindent where $V_i$ is the amplitude of the RF voltage applied to the $i^{th}$ electrode, $J_1$ is the Bessel function of first order, and the $r_i$ are the radii describing the trap geometry (as given in \fig{fig1}). There is a simple expression for the potential due to a given electrode carrying a voltage $V$ along the trap axis (at $r=0$), given by 

\begin{equation}
\Phi(z) = V \left[ \left(1+\left(\frac{r_1}{z}\right)^2 \right)^{-1/2} - \left(1+\left(\frac{r_2}{z}\right)^2 \right)^{-1/2} \right],  
\end{equation} 

\noindent where $r_2$ and $r_1$ are the inner and outer radii, respectively, for that particular electrode \cite{THKim:10}. The full potential may be obtained by adding all potentials due to each RF electrode. In the work that follows, we focus on motion in the $z$ direction; therefore, we will rely heavily on this expression. 

% Add here information on the actual shape of the trap. 

\begin{figure}
\begin{center}
\includegraphics[width=.6\textwidth]{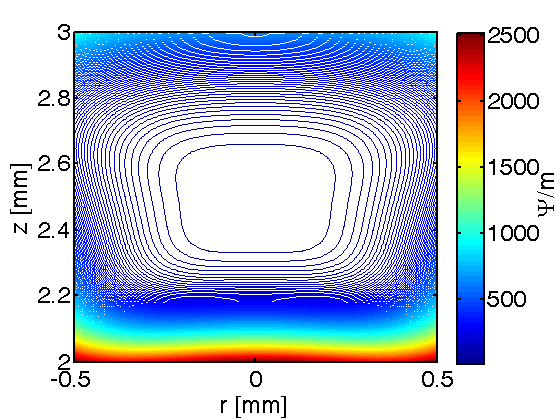}
\end{center}
\caption{Cross-section of the pseudopotential. The pseudopotential $\Psi$ (in units of eV) divided by the mass $m$ (in units of kg) is plotted, so that the charge to mass ratio $R$ found experimentally may be employed. }\label{fig:fig2}
\end{figure}

\begin{figure}
\begin{center}
\includegraphics[width=.6\textwidth]{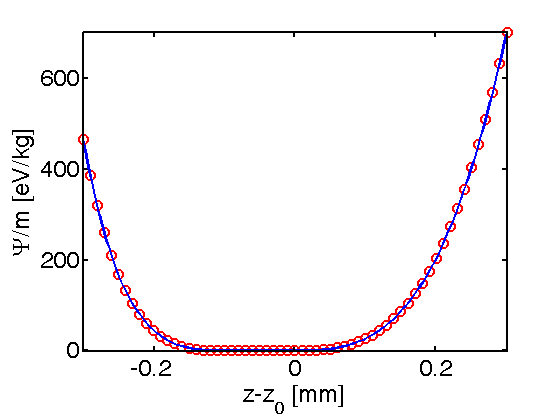}
\end{center}
\caption{The pseudopotential (divided by mass) along the $z$ direction at $r=0$ (along the trap axis). A sixth-order fit is made to this potential. Writing the potential as $\Psi/m(z-z_0)=\sum_{n=0}^6 c_n (z-z_0)^n$, the fit parameters are $c_6=9.87\times 10^4$~eV/kg/mm$^6$, $c_5=-1.11\times 10^5$~eV/kg/mm$^5$, $c_4 = 5.37 \times 10^4$~eV/kg/mm$^4$, $c_3 = 1.44 \times 10^4$~eV/kg/mm$^3$, $c_2 = 7.79\times 10^2$~eV/kg/mm$^2$, $c_1 = -2.04$~eV/kg/mm, and $c_0 = -0.015$~eV/kg. The range for this plot and fit is chosen to comfortably accommodate the range of motion of a trapped particle in our experiments.  }\label{fig:zfit}
\end{figure}

We now present the results of our modeling of the trap. As in the experiments described herein, we assume that the RF voltage applied to electrodes RF1 and RF3 has an amplitude of 660~V at a frequency of 297~Hz. All other electrodes are assumed to be at RF ground. We use the experimentally-obtained value of $R \equiv  Q/m \approx -2.5\times 10^{-3}$~C/kg for the charge-to-mass ratio of a particular ion. A two-dimensional cross-section of $\Psi(r,z)$ is shown in \fig{fig2}. In \fig{zfit}, we plot $\Psi(r=0,z)/m$ and a fit thereof, in order to show the importance of the various higher-order terms. 

\begin{figure}
\begin{center}
\includegraphics[width=.6\textwidth]{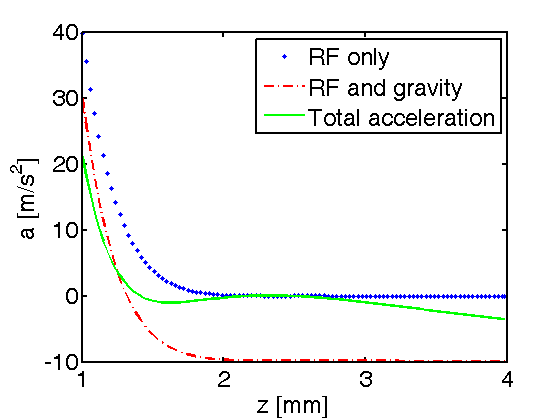}
\end{center}
\caption{Plot of the acceleration experienced by a trapped ion as a function of position. The dotted (blue online) curve only accounts for the pseudo-force from the RF electrodes. The dash-dot (red online) curve shows the offset due to gravity; it is merely the blue curve translated down by 9.8~m/$\mathrm{s}^2$. The solid (green online) curve is the total acceleration, including the effect of the DC voltage of -37~V applied to RF2. The ion indeed is uncompensated (located below the point at which the RF electric field vanishes) when it is at the point at which the acceleration vanishes. However, when it is excited by an applied oscillating voltage, it is very nearly compensated at its maximum value of $z$. The total acceleration has a limiting value of -9.8~m/s$^2$ as $z \rightarrow \infty$.}\label{fig:fig3}
\end{figure}

\begin{figure}
\begin{center}
\includegraphics[width=.6\textwidth]{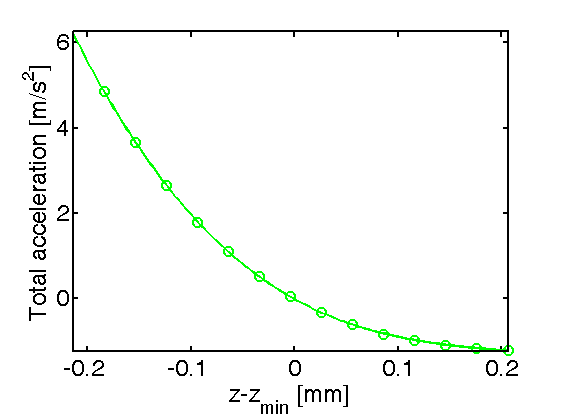}
\end{center}
\caption{Plot of the total acceleration of a trapped ion, zoomed in on the true equilibrium position of the ion at $z_{min} = 1.4$~mm. Although the potential has been altered by the need to compensate for the effect of gravity on the ion, the acceleration is still substantially nonlinear. A third-order polynomial fit is the solid line intersecting the computed data points. Writing the acceleration as $a(z-z_{min})=\sum_{n=1}^3 d_n (z-z_{min})^n$, the fit coefficients are $d_3 = -87.7$~m/s$^2$/mm$^3$, $d_2 = 54.8$~m/s$^2$/mm$^2$, and $d_1 = -13.4$~m/s$^2$/mm.}\label{fig:totala_fit}
\end{figure}

In experiments with atomic ions, the gravitational force on the particle can normally be neglected in calculating the pseudopotential. However, in the case of trapping micron-scale objects, the force of gravity is not negligible. This is especially so given the flat landscape of the RF potential near the trap center. In order to compensate the trap (which means to null the micromotion of the ion), a DC voltage $V_{\mathrm{DC}}$ is applied to the electrode RF2. This DC voltage pushes the ion toward the point at which the RF electric field vanishes. For our present work, a value of $V_{\mathrm{DC}}=-37$~V was used. Since the resulting DC electric field varies spatially, an accurate model of the trap must include the static potential $\Phi_{\mathrm{DC}}(r,z)$ which has the charge on electrode RF2 as its source. We use an effective force model to describe the trap. We first define the effective pseudo-force as $F_{\mathrm{RF}} = -\nabla \Psi(r,z)$. In the following analysis, we consider only the $z$ direction, as the ion is confined approximately to the z-axis. (We say ``approximately'' since, in our experiments, the amplitude of motion in the $r$ direction is on the order of one-tenth the amplitude in the $z$ direction.) Considering this, the pseudo-force is written as $F_{\mathrm{RF}} = -\partial \Psi(z)/ \partial z$, where $\Psi(z) \equiv \Psi(r=0,z)$. The corresponding force from the applied DC compensation voltage is then $F_{\mathrm{DC}} = - Q \partial \Phi_{\mathrm{DC}}(z) / \partial z$. The total force is then given by 

\begin{equation} 
F = F_{\mathrm{RF}} + F_{\mathrm{DC}} -mg, 
\end{equation} 

\noindent where $g$ is the acceleration due to gravity. So that this expression may be written in terms of the charge-to-mass ratio $R$ of the ion, it is convenient to recast it into an expression for the acceleration $a(z)$ of the ion. In so doing, we obtain 

\begin{equation} 
a(z) = -\frac{R^2}{4 \Omega^2} \frac{\partial}{\partial z}\left | E(z)\right |^2 + R E_{\mathrm{DC}} - g,  
\end{equation} 

\noindent where $E(z)$ is the oscillating electric field due to the potentials on RF1 and RF3, and $E_{\mathrm{DC}}$ is the static electric field due to the potential on RF2. In \fig{fig3} and its caption we demonstrate the effects of gravity and of the applied DC voltage on the acceleration experienced by a trapped ion. \fig{totala_fit} shows the acceleration experienced by a trapped ion near its equilibrium position, and a fit of that function is given. 

\section{The Experiment}

For this initial experiment with multipole traps, we use micron-sized granules of sugar which are held in air (inside an airtight container). This is a suitable situation for this first set of measurements on SEMT's since it is much simpler than a comparable setup for atomic ions. However, it carries the disadvantage that the charge and mass of the ion are not known in advance. Therefore when we seek to match our experimental data with a model, we must adjust the model according to various charge and mass values until the model converges to the experimental data.  

The trap is loaded by tapping confectioner's sugar from a plastic spatula onto the trap. At times, we use compressed air to blow the sugar off the trap surface. Normally, several particles are trapped and we cull the number to just one by manually adjusting the DC voltages. A single grain of sugar can be stored for several hours; typically, it is lost only when we shut the trap down. The ion scatters red laser light from a beam with a power of $\approx 2$~mW and a width of $\approx 1$~mm. The light is collected using an imaging system with resolution 12~$\mu$m and magnification 3.4, which is calibrated using a USAF 1951 resolution target. The scattered light is recorded using a Thorlabs DCC3240C CMOS camera. 

We measure the trajectory of the ion as a function of time that results from an applied low-frequency signal on one of the DC electrodes. At these low frequencies, we simply use an op-amp adder circuit to combine the DC offset with the oscillating voltage, called the ``tickle.'' This combined signal is sent to an amplifier (Piezo Systems EPA-008-1) which has a gain of 20 in our frequency range. The use of an op-amp circuit has a nice side-benefit: the rails are fixed at +7~V and -7~V, which limits the input voltage to the amplifier; its maximum safe input is $\pm 9$~V. Tickle frequencies in the range of 1 to 50~Hz may be used, depending on the framerate of the camera. The maximum framerate depends on the size of the window, which itself depends on the amplitude of the ion's motion in the trap. Typically, a window may be selected such that the framerate is approximately 60~$s^{-1}$, which gives a Nyquist frequency of 30~Hz. Since a primary focus of the multipole trap measurements is frequency doubling, a nominal maximum tickle frequency is 15~Hz. The data presented below were obtained using a measured tickle amplitude of 10.4~V. 

To obtain the data presented here, we record videos in avi format, typically consisting of 500 individual frames. These videos are converted into a set of images using the software ffmpeg \footnote{https://www.ffmpeg.org/}, which is called from Matlab, which is used to process the data. These images are then imported into Matlab, and the center of each ion is found. We use an algorithm which effectively calculates the ``center of mass'' for a given image, using the brightness of each pixel. These data in sequence form the position vs. time trajectory of the ion. Our method is very similar to that employed in the work of Ref.~\cite{Madsen:14}, of which we became aware while our experiments were underway. 

Understanding the data requires comparing it with the model of the trap described above. This requires integrating the equations of motion for a single ion of a given $R$, then attempting to adjust $R$ until a fit to the observed data is found. There is an additional term that depends on the mass but not on the charge, which must be included in order to get an accurate simulation: the damping force from air. This force, denoted $F_{\mathrm{D}}$, is described by

\begin{equation}\label{eq:damping}
F_{\mathrm{D}} = - \mu v, 
\end{equation}

\noindent where the constant $\mu$ quantifies the level of damping and $v$ is the velocity of the particle in the $z$ direction. There are, in total, two parameters that must be fit, $R$ and $\mu$. If the ion were spherical in shape, we could use the result quoted in Ref.~\cite{Pearson:06}, which states that 

\begin{equation} \label{eq:mumu}
\mu = 6\pi \mu_D r_p, 
\end{equation} 

\noindent where $\mu_{\mathrm{D}} = 1.83\times 10^{-5}$~kg/m$\cdot$s is the dynamic viscosity of air and $r_p$ is the radius of the particle. However, we lack the optical resolution to determine the shape of the particle; therefore we simply find the value of $\mu$. 

% Have to explain tickle here before showing data 
\begin{figure}
\begin{center}
\includegraphics[width=.6\textwidth]{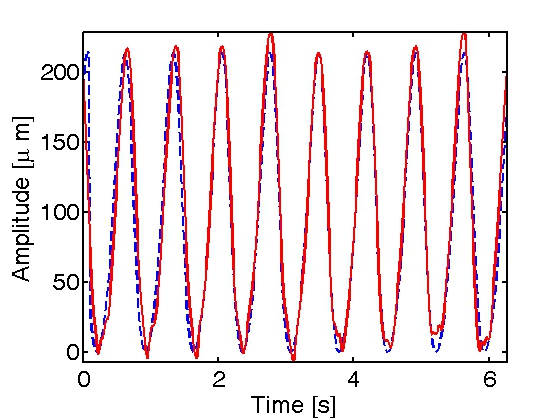}
\end{center}
\caption{Simulated (blue dashed line) and measured (red solid line) trajectory of an ion being excited at a frequency of 1.4~Hz by an oscillating voltage with amplitude 10.4~V applied to the electrode RF2. Our result values of $R = -2.54$~mC/kg and $\mu / m=674$~s$^{-1}$ are used in the simulation. The average height of the ion above the trap surface during the experiment is 1.4~mm (as shown in \fig{fig3} and \fig{totala_fit}).}\label{fig:fig4}
\end{figure}

\begin{figure}
\begin{center}
\includegraphics[width=.6\textwidth]{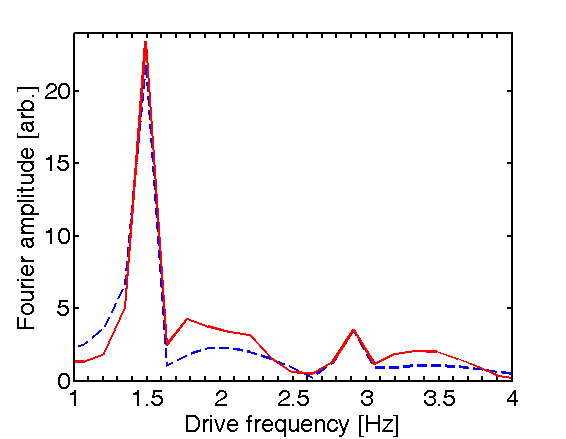}
\end{center}
\caption{Fourier transform of the trajectory shown in \fig{fig4}. The simulation is the blue dashed line, and the experimental data is the red solid line. }\label{fig:fig5}
\end{figure}

\begin{figure}
\begin{center}
\includegraphics[width=.6\textwidth]{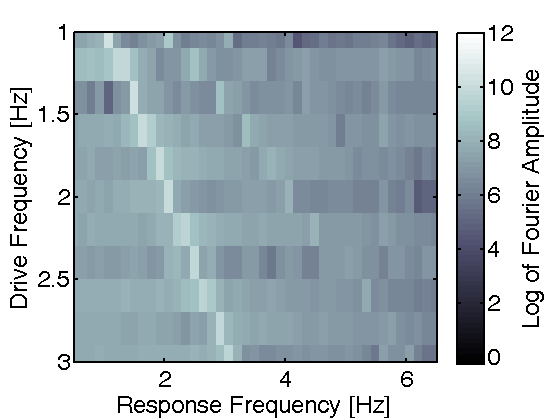}
\end{center}
\caption{Two dimensional plot of the frequencies at which the ion responds versus the tickle frequency. The false-color plot is of the logarithm of the Fourier amplitude at each pair of frequencies. Frequency doubling is seen up to a frequency of about 6~Hz, after which it is difficult to see on this plot. One reason for this suppression of frequency doubling is the air damping, the force from which is velocity-dependent.}\label{fig:fig6}
\end{figure}

In \fig{fig4}, we present one example trajectory, for tickle frequency 1.4~Hz. The ion is clearly responding to the sinusoidal tickle according to a nonsinusoidal waveform. The nonlinear properties of the trap may be most readily seen by Fourier transforming the position-vs-time data, revealing the frequencies present in the trajectory. We find a strong peak at double the tickle frequency and a lesser peak at triple the frequency. An example of this Fourier transformed data is presented in \fig{fig5}, and \fig{fig6} is a two-dimensional plot of these data for several frequencies. 

\begin{figure}[!htbp]
\begin{center}
\includegraphics[width=.6\textwidth]{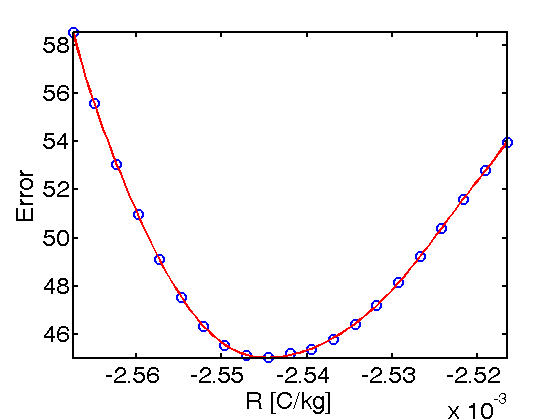}
\end{center}
\caption{A typical plot of the error between the simulation and experiment as a function of simulated charge to mass ratio $R$. The error is calculated as the rms difference between the simulation and the experiment; therefore it has units of $\mu$m. The red line connecting the data points is a fit of an eighth-order polynomial to the data. An eighth-order polynomial is used in our fitting routine so that wider ranges of $R$ can be tested, but the value of $R$ here does not differ substantially from that obtained by a quadratic fit. The minimum value of $R$ is obtained by finding a local minimum of that fit function within the range of simulated $R$ values used.}\label{fig:fig7}
\end{figure}

For each trajectory, recorded at a different tickle frequency, we obtain a value for the charge-to-mass ratio $R$. The rms difference between the computed and measured trajectory is the error for each simulation, and the value of $R$ which minimizes the error is found. A sample plot of this error as a function of simulated $R$ is given in \fig{fig7}. For the particular ion used when the data were taken, we obtain a value of $R = -2.54 \pm 0.01$~mC/kg. This is a 1-$\sigma$ statistical error bar. There are, however, good reasons to believe that systematics dominate the error in the measurement, not the least of which is that important parameters such as the imaging system magnification and RF voltage are known to only two significant digits. We estimate the systematic error on $R$ in the next section. Our measured value of $R$ is well in line with previous measurements on charged dust particles \cite{Madsen:14}. The statistical error on the damping coefficient $\mu$ is greater because it affects the trajectory far less than $R$. For our result value of $R$, we obtain $\mu/m = 679 \pm 46$~kg/s. (Here, $m$ is the particle mass.) This value is found in a similar way to the value for $R$, by minimizing the error between the experimental and simulated trajectories. The measurement of $\mu/m$ permits us to estimate the size of the trapped particles. They are almost certainly not spherical in shape, but if they were, we could invoke \eq{mumu} and use $\rho_{s} (\mu/m)(4\pi/3) r_p^3 = 6\pi \mu_D r_p$, where $\rho_s = 1590$~kg/m$^3$ is the density of sucrose, to obtain a very reasonable value of $r_p \approx 9$~$\mu$m. This results in a particle mass $m$ of about 5~fg, and a charge on the particle of $Q = mR = 1\times 10^{-14}$~C. These values are only estimates, but it appears in principle, for a particle whose drag is well-understood, to be possible to ascertain both the charge and the mass using our methods, not merely the ratio $R$. 

\section{Discussion}

In this article, we have shown evidence that an ion trapped in our SEMT exhibits a nonlinear response to an applied signal. Although this has been seen in ion traps before, this experiment is unique in that the ion is close to the center of the trap during the experiment. The ion is actually close to being perfectly compensated when it is at its maximum height; we determine this by noting that the extent to which it is undergoing micromotion in the vertical direction is minimized at that point. Naturally, it would be ideal to repeat this experiment in vacuum, with an atomic ion of known charge and mass. One reason is that the effect of gravity on the atomic ion is negligible, and so a large compensation for the weight of the particle using an electric field sourced by RF2 would not be required, and we would therefore be able to see the true behavior of the multipole trap, without the quadratic terms introduced by RF2. Another aspect that would improve is that the damping of an atomic ion due to the laser cooling force would be easier to model than damping due to air, and we could therefore see a more direct confrontation between the experiment and the model. Nevertheless, with the experiment reported here, we see a clear qualitative signature of nonlinear dynamics, and find that our model yields a very reasonable value for the charge to mass ratio ($R$) of the ion. We have also contrasted the observed multipole trap behavior with quadrupole trap behavior by trapping ions in quadrupole configuration (by applying the RF trapping voltage to RF2 while grounding RF1 and RF3). We see no frequency upconversion in this case; the ion trajectory simply follows the drive. We also see the appearance of a secular frequency (the hallmark of a harmonic trap) in the Fourier transform of the trajectories. However, due to the presence of air, we do not see an increased amplitude of motion when the ion is driven at the secular frequency. Simulations verify that no resonant excitation should be observable, for our ion properties and other parameters. Future experiments in which we pump the system down should allow us to observe resonant excitation. As it stands, the shape of the potential itself depends strongly on the properties of the ion; therefore we cannot presently do a quantitative measurement of the size of each term in the potential. That said, there is no reason to doubt that the electrostatics used in the trap modeling are reliable. 

Since several of our experimental parameters are known to only two significant digits, we would expect systematic error on the measurement of $R$ on the order of a few percent. We estimate the systematic error in the following way. We note that the \emph{shape} of the trajectory alone should be sufficient to calculate $R$, without regard to the absolute amplitude of the trajectory. This is because the trap acceleration function is both non-linear and non-periodic; there is only one position along the curve that can generate a particular response trajectory from the ion. Therefore, we repeat the fitting process, but normalizing the trajectories each time (to have a maximum value 0.5 and minimum value -0.5). For each dataset, the value of $R$ that minimizes the error is found. Using this approach, we obtain $R = -2.69\pm 0.04$~mC/kg. This is an error of 6\% compared with the other measurement, in line with our expectations. We have confined a few other sugar particles in our trap, and done a rough estimate of their $R$ values using the methods outlined in this paper. The mean value is around 7 mC/kg with a standard deviation (over six measurements) of 5 mC/kg. This ought not be surprising, since they are charged triboelectrically. 

Since these measurements represent a new way (to our knowledge) of determining $R$ for some substance, it is natural to ask whether they may be adapted into a new type of mass spectrometry. Perhaps, since only the shape of the trajectory needs to be known (rather than the absolute amplitude), the excitation of the motion of the unknown ion could be detected electronically. Furthermore, SEMT's have several other desirable properties. SEMT's, in contrast to quadrupole traps, appear to be able to simultaneously confine particles of widely different $R$ values. In addition, the depth of the trap, as commonly calculated for quadrupole traps, is significantly greater than that of quadrupole surface traps of similar scale \cite{Clark:13}. This leads us to the idea that an array of SEMT's could work as a  mass analyzer, providing a broad spectrum spanning many $R$ values simultaneously. The traps would need to be very small, and multiple traps would need to be operated in parallel, but that is established technology; a microfabricated array of quadrupole ion traps for mass spectrometry has already been demonstrated \cite{Pau:06}. Further work on optimization of the trap geometry, as well as mapping of the stability diagram of the traps, will help to resolve whether this is a possibility. Successful operation would require marked improvement in systematic sources of error, which would likely be possible when trapping atomic or molecular ions, especially if the trap can be calibrated using a particle of known $R$. 

We gratefully acknowledge funding from the Research Corporation for Science Advancement under a Cottrell College Science Award, and also funding from the Citadel Foundation.

\bibliographystyle{apsrev}
%\bibliography{refs_master}

\end{document}